\documentclass[11pt]{article}
\usepackage{type1cm,amsmath,hangcaption,graphicx,indentfirst}

\begin{document}



\begin{center}
{\bf \Large \tt \Large A Survey of Finite Algebraic Geometrical
Structures Underlying Mutually Unbiased Quantum Measurements}
\end{center}


\bigskip

\begin{center} {\bf Michel Planat\footnote{\noindent Institut FEMTO-ST, Departement LPMO, 32 Avenue de l'Observatoire,

$\quad$25044 Besan\c con Cedex, France; e-mail: planat@lpmo.edu},
Haret C. Rosu\footnote{\noindent Potosinian Institute of Science and
Technology (IPICyT), Apdo Postal 3-74

$\quad$Tangamanga, 78231 San Luis Potos\'{\i}, S.L.P., Mexico;
e-mail: hcr@ipicyt.edu.mx}, Serge Perrine\footnote{\noindent France
Telecom, Conseil Scientifique, 38-40 rue du G\'en\'eral
Leclerc,

$\quad$92794 Issy les Moulineaux Cedex 9, France; e-mail:
serge.perrine@wanadoo.fr}

} \end{center}

\bigskip
\bigskip
\bigskip

\noindent The basic methods of constructing the sets of mutually
unbiased bases in the Hilbert space of an arbitrary finite dimension
are reviewed and an emerging link between them is outlined. It is
shown that these methods employ a wide range of important
mathematical concepts like, e.g., Fourier transforms, Galois fields
and rings, finite and related projective geometries, and
entanglement, to mention a few. Some applications of the theory to
quantum information tasks are also mentioned.

\bigskip




\section{INTRODUCTION}
\label{Intro}

Problems pertinent to quantum information theory are touching more
and more branches of pure mathematics, such as number theory,
abstract algebra and projective geometry. This paper focuses on one
of the most prominent issues in this respect, namely the
construction of sets of mutually unbiased bases (MUBs) in a Hilbert
space of finite dimension. For an updated list of open problems
related to the development of quantum technologies the reader is
directed to the Quiprocon website \cite{Quiprocone}.

To begin with, we recall that two different orthonormal bases $A$
and $B$ of a $d$-dimensional Hilbert space $\cal{H}$$^{d}$ with
metrics $\langle \ldots|\ldots\rangle$ are called mutually unbiased
if and only if $|\langle a|b\rangle|= 1/\sqrt{d}$ for all
$a$$\in$$A$ and all $b$$\in$$B$. An aggregate of MUBs is a set of
orthonormal bases which are pairwise mutually unbiased. The MUBs
have been first studied by Schwinger in 1960 \cite{s60}. Two decades
later, important results by Alltop \cite{a80} passed unnoticed and
even a well-published paper by Ivanovi\'c \cite{iv81} still did not
trigger their systematic research although he proved  that the
maximum number of such bases is $d+1$ when $d$ is a prime. The
latter began with the important paper of Wootters and Fields
\cite{Wootters89} in which they showed how to construct the maximum
number of such bases, i.e. $d+1$, for $d$ a power of a prime. Yet, a
still unanswered question is if there are non-prime-power values of
$d$ for which this bound is attained. It has been surmised
\cite{Archer} \cite{Klapp03} that the maximum number of such bases,
$N(d)$, is equal to $1+ min(p_i^{e_i})$, the latter quantity being
the lowest factor in the prime number decomposition of $d$,
$d=\prod_ip_i^{e_i}$ (for a violation of this bound, see the recent
work of Wocjan and Beth \cite{Wocjan04} and/or our comment at the
end of Section 4 herein). For example, it is still not known
\cite{Grassl} whether there are more than three MUBs for $d=6$, the
lowest non-prime-power dimension, although the latest findings of
Wootters \cite{Wootters04} (and an earlier result of G. Tarry quoted
in the last reference) seem to speak in favor of this conjecture.
Klappenecker and R\"otteler \cite{Klapp03} showed that at least 3
MUBs exist in any dimension 
and some conditions for the existence of more than 3 MUBs for any
dimension are also known \cite{comb06}.

MUBs have already been recognized to play an important role in
quantum information theory. Their main domain of applications is the
field of secure quantum key exchange (quantum cryptography). This is
because any attempt by an eavesdropper to distinguish between two
non-orthogonal quantum states shared by two remote parties will
occur at the price of introducing a disturbance into the signal,
thus revealing the attack and allowing to reject the corrupted
quantum data. Until recently, most quantum cryptography protocols
have solely relied, like the original BB84 one, upon 1-qubit
technologies, i.e., on the lowest non-trivial dimension ($d=2$),
usually the polarization states of a single photon, or other schemes
such as the sidebands of phase-modulated light \cite{Merolla99}. But
security against eavesdropping has lately been found to
substantially increase by using all the three bases of qubits,
employing higher dimensional states, e.g. qudits
\cite{Nielsen00},\cite{Cerf01}, or even entanglement-based protocols
\cite{Durt03}. Another, closely related, application of MUBs is the
so-called quantum state tomography, which is thought to be the most
efficient way to decipher an unknown quantum state
\cite{Quiprocone}.

Quantum state recovery and secure quantum key distribution can also
be furnished in terms of so-called positive operator valued measures
(POVMs) which are symmetric informationally complete (SIC-POVMs)
\cite{Renes03}. These are defined as sets of $d^2$ normalized
vectors $a$ and $b$ such that $|\langle a|b\rangle|= 1/\sqrt{d+1}$,
where $a \neq b$, and they are connected with MUBs. Unlike the
latter ones, however, the SIC-POVMs are likely to exist in all
finite dimensions and they have already been constructed for $d=6$
\cite{Grassl}. The intricate link between MUBs and SIC-POVMs has
recently been examined by Wootters \cite{Wootters04} and acquired an
intriguing geometrical footing in the light of the ``SPR conjecture"
\cite{Saniga} stating that the question of the existence of a set of
$d + 1$ MUBs in a $d$-dimensional Hilbert space if $d$ differs from
a power of a prime number is equivalent to the problem of whether
there exist projective planes whose order $d$ is not a power of a
prime number. Also, Bengtsson and Ericsson \cite{be05} provided a
connection with sets of $d^2$ facets of convex polytopes of power of
prime dimensions: the centers of the facets can form a regular
simplex if and only if there is an affine plane of order $d$ that
exists only if $d$ is a power of a prime.

We also mention the interesting fact recently noticed in the Lie
algebra approach to MUBs \cite{boy05} that a complete collection of
MUBs in $C^d$ gives rise to a so-called orthogonal decomposition
(OD) of $sl_d(C)$ for which a longstanding conjecture says that ODs
of $sl_d(C)$ can only exist if $d$ is a prime power. On the other
hand, for recent works on the relationship between MUBs and the
SU(2) theory of quantum angular momentum the reader is directed to
\cite{Kib1}\cite{Kib2}(in \cite{Kib2} a compact formula for MUBs is
 given in the case where $d$ is a prime number).

This survey is organized as follows. In Sections
\ref{MUBGaloisFields} and \ref{MUBsQubits} the construction of a
maximal set of MUBs in dimension $d=p^m$, $p$ being a prime, as a
quantum Fourier transform acting on a Galois field ($p$ odd) and a
Galois ring $GR(4^m)$ ($p=2$) is discussed. This puts in perspective
the earlier formulas by \cite{Wootters89} and \cite{Klapp03},
respectively. The case of non-prime-power dimensions is briefly
examined in Section \ref{compos}. Next, in Section \ref{projective},
we focus on our recent conjecture on the equivalence of two
problems: the surmised nonexistence of projective planes whose order
is not a power of a prime and the suspected non existence of a
complete set of MUBs in Hilbert spaces of non-prime-power
dimensions. The geometry of qubits is discussed and the concept of a
lifted Fano plane is introduced. Finally, an intricate relationship
between MUBs and maximal entanglement is emphasized in Section
\ref{seis}, which promises to shed fresh light on newly emerging
concepts such as the distillation of mixed states and bound
entanglement \cite{hor3}. The entanglement properties of MUBs for
systems of three and four qubits have been recently discussed in
detail by Romero and collaborators \cite{rom05}. We endeavored to
make the paper as self-contained as possible from our standpoint.
Yet, the interested reader may find it helpful to consult some
introductory texts on quantum theory in a finite Hilbert space and
its relation to Fourier transforms and phase space methods, e.g.,
the reviews by A. Vourdas \cite{Vourdas04}.

\section{MUB'S, QUANTUM FOURIER TRANSFORMS AND GALOIS FIELDS} 
\label{MUBGaloisFields}

In order to see the close connection between MUBs and Fourier
transforms, we consider an orthogonal computational basis
\begin{equation}
B_0=(|0\rangle,|1\rangle,\cdots,|n\rangle, \cdots, |d-1\rangle)
\end{equation}
with indices $n$ in the ring $\mathcal{Z}_d$ of integers modulo $d$.
There is a dual basis which is defined by the quantum Fourier
transform
\begin{equation}
|\theta_k\rangle=\frac{1}{\sqrt{d}}\sum_{n=0}^{d-1}\omega_d^{k
n}|n\rangle, \label{Fourier}
\end{equation}
where $k \in \mathcal{Z}_d,  \omega_d=\exp(\frac{2 i \pi}{d})$ and
$i^{2}=-1$.

\noindent In the context of quantum optics this Fourier transform
relates Fock states $|k\rangle$ of light to the so-called phase
states $|\theta_k\rangle$. The properties of the quantum phase
operator underlying this construction have extensively been studied
and found to be linked to prime number theory \cite{Planat04}.

\subsection{${\bf d=2}$: The quantum gates approach}

For $d=2$, i.e. the case of qubits, one has $\omega=-1$ and so
\begin{equation}
|\theta_0\rangle=\frac{1}{\sqrt{2}}(|0\rangle+|1\rangle);~~
|\theta_1\rangle=\frac{1}{\sqrt{2}}(|0\rangle-|1\rangle).
\end{equation}
These two vectors can also be obtained by applying the Hadamard
matrix $H=\frac{1}{\sqrt{2}}\left [\begin{array}{cc} 1 & 1\\ 1 & -1
\end{array}\right]$ to the basis $(|0\rangle,|1\rangle)$. Note that the two orthogonal bases
$B_0=(|0\rangle,|1\rangle)$ and
$B_1=(|\theta_0\rangle,|\theta_1\rangle)$ are mutually unbiased. The
third basis $B_2=(|\psi_0\rangle,|\psi_1\rangle)$ which is mutually
unbiased to both $B_0$ and $B_1$ is obtained from $H$ by the
pre-action of a $\pi/2$ rotation $S=\left[\begin{array}{cc} 1 & 0\\
0
&i\end{array}\right]$, so that $HS=\frac{1}{\sqrt{2}}\left[\begin{array}{cc} 1 & i\\
1 & -i\end{array}\right]$. The three matrices $(I,H,HS)$ thus
generate the three MUBs. These matrices are also important for two
qubits gates in quantum computation \cite{Nielsen00}.

\subsection{${\bf d=2}$: The Pauli matrices approach}

The above-outlined strategy for finding MUBs for qubits contrasts
with that used by the majority of authors. The eigenvectors of the
Pauli spin matrices
\begin{equation}
\sigma_z=\left [\begin{array}{cc} 1 & 0\\
0 & -1\end{array}\right],  \sigma_x=\left [\begin{array}{cc} 0 & 1\\
1 & 0
\end{array}\right], \sigma_y=\left [\begin{array}{cc} 0 & -i\\ i & 0
\end{array}\right]~,
\end{equation}
where $\sigma_y=i\sigma_x \sigma_z$, are precisely the sought bases
$B_0$, $B_1$ and $B_2$.

\noindent A natural generalization of Pauli operators $\sigma_x$ and
$\sigma_z$ for an arbitrary dimension $d$ is the Pauli group of
shift and clock operators:
\begin{eqnarray}
&X_d|n\rangle=|n+1\rangle,\\ \nonumber
&Z_d|n\rangle=\omega_d^n|n\rangle. \label{Pauli}
\end{eqnarray}
For a prime dimension $d=p$, it can be shown that the eigenvectors
of the unitary operators $(Z_p,X_p,X_pZ_p,\cdots,X_pZ_p^{p-1})$
generate the set of $d+1$ MUBs \cite{Bandyo01}. A natural question
here emerges whether this method can straightforwardly be
generalized to any dimension.


\subsection{MUBs on Galois fields of odd characteristic}

Let us attempt to rewrite Eq.~(\ref{Fourier}) in such a way that the
exponent of $\omega_d$ now acts on the elements of a Galois field
$G=GF(p)$, the finite field of integers modulo an odd prime $p$ (or
finite field of odd prime characteristic $p$) and cardinality $d=p$,
which in general are defined by an irreducible polynomial (see Sect.
3). Denoting ``$\oplus$" and ``$\odot$" the two usual operations in
the field and replacing $\omega_d$ by the root of unity $\omega_p$,
we get
\begin{equation}
|\theta_k\rangle=\frac{1}{\sqrt{d}}\sum_{n=0}^{d-1}\omega_p^{k\odot
n}|n\rangle. \label{Galois}
\end{equation}
Next, we employ the Euclidean division theorem for fields
\cite{Lidl}, which says that given any two polynomials  $k$ and $n$
in $G$ there exists a uniquely determined pair $a$ and $b$ in $G$
such that $k=a \odot n \oplus b$, $\deg{b}<\deg{a}$. This allows for
the exponent in Eq.~(\ref{Galois}), $E$, to be written as $E=(a
\odot n \oplus b)\odot n$. In the case of prime dimension $d=p$, $E$
is an integer and the sum in Eq.~(\ref{Galois}) is well defined.

For $G=GF(p^m)$, the field is defined by an irreducible polynomial,
the cardinality is $d=p^m$ and $E$ is a polynomial too. In this
case, instead of Eq.~(\ref{Galois}) one should use the following
expression
\begin{equation}
|\theta_b^a\rangle=\frac{1}{\sqrt{d}}\sum_{n=0}^{d-1}\omega_p^{tr[(a
\odot n \oplus b)\odot n ]}|n\rangle, \label{newGalois}
\end{equation}
where ``$tr$" stands for the trace of $GF(p^m)$ down to $GF(p)$,
\begin{equation}
tr(E)=E\oplus E^p \oplus \cdots \oplus E^{p^{m-1}},~~E \in GF(p^m).
\label{trace}
\end{equation}

If $p$ is odd, Eq.~(\ref{newGalois}) defines the set of $d$ bases,
with the index $a$ for the basis and the index $b$ for the vector in
the bases mutually unbiased to each other and to the computational
basis $B_0$ as well. In a slightly different form, this equation was
first derived by Wootters and Fields \cite{Wootters89}. Its nice
short elucidation, based on Weil sums, is due to Klappenecker and
R\"otteler \cite{Klapp03}. Another, a more tricky derivation still
in the spirit of Fourier transforms and claimed to hold also for the
case of characteristic $2$, was found by Durt \cite{Durt04}. An
interesting approach based on the Weyl operators in the $L^2$-space
over Galois fields is also worth mentioning \cite{Partha04}.

As already pointed out by Wootters and Fields \cite{Wootters89}, the
reason why (\ref{newGalois}) defines the complete set of MUBs relies
on the field theoretical formula $|\sum_{n=0}^{d-1}\omega_p^{tr[(a
\odot n \oplus b)\odot n ]}|n\rangle|=p^{1/2}$, with $ a \neq 0$ and
$p$ being an odd prime. This method, however, fails for
characteristic two where $|\sum_{n=0}^{d-1}\omega_2^{tr[(a \odot n
\oplus b)\odot n ]}|n\rangle|=0$ for any $a$, $b$. As shown in Sect.
3 below, for this characteristic one has to use Galois rings instead
of Galois fields to get a complete set of MUBs.

A closer inspection of (\ref{newGalois}) reveals an intricate
relation between MUBs and quantum phase operators. It is known
\cite{Planat04} that the Fourier basis $|\theta_k\rangle$ can be
derived in terms of the eigenvectors of a quantum phase operator
with eigenvalues $\theta_k$ and given by
$\Theta_d=\sum_{k=0}^{d-1}\theta_k|\theta_k\rangle\langle
\theta_k|$. Similarly, using well known properties of the field
trace, one can show that each base of index $a$ can be associated
with a quantum phase operator
\begin{equation}
\Theta_d^a=\sum_{b=0}^{d-1}\theta_b^a|\theta_b^a\rangle\langle\theta_b^a|,
\label{quantumphaseop}
\end{equation}
with eigenvectors $|\theta_b^a\rangle$ and eigenvalues $\theta_b^a$;
the latter may thus be called an ``MUB operator".

\section{MUBs FOR EVEN CHARACTERISTIC FROM GALOIS RINGS} 
\label{MUBsQubits}

Our next goal is to find a Fourier transform formulation of MUBs in
characteristic $2$.
This is a very important case since the `$2^m-$dits' are the basic
information units in quantum computation. One may be tempted to
connect the Galois field algebra with the generalized Pauli
operators (\ref{Pauli}) by constructing discrete vector spaces over
the Galois field \cite{Wootters04bis}. For the one qubit case we
already know that the eigenvectors of Pauli matrices $\sigma_z$,
$\sigma_x$ and $\sigma_x \sigma_z$ define the three MUBs. Passing to
the quartit (i.e., $4$-dit) case, one finds that the operators of
the following tensorial products $\sigma_z\otimes \sigma_x$,
$\sigma_z \otimes \sigma_x \sigma_z$ and $\sigma_x\sigma_x \otimes
\sigma_z$ are associated to translations, i.e., to a single
one-dimensional subspace in the corresponding vector space, and they
define a unique basis represented by their simultaneous
eigenvectors. Since there are $4+1$ one-dimensional subspaces in
this discrete vector space, there are also $4+1$ MUBs. Other
geometrically inspired derivations based on the tensorial
decomposition of operators in the Pauli group can be found in the
literature \cite{Bandyo01}\cite{Mosseri03}\cite{Pittenger03}.

On the other hand, we are interested in a result equivalent to the
formulas given in Eqs.~(\ref{Galois}) and (\ref{newGalois}) for the
case of characteristic two. Instead of the Euclidean division in the
field $GF(2^m)$, it is necessary to consider a decomposition in the
Galois ring $GR(4^m)$ (defined below) so that the relevant root of
unity in the Fourier formula now reads $\omega_4=\exp(2i\pi/4)=i$.
For qubits $GR(4)=\mathcal{Z}_4$, and in general any number $k$ in
$\mathcal{Z}_4$ can be written as $k=a\oplus 2 \odot b$,
where $\oplus$ and $\odot$ now act in $\mathcal{Z}_4$. 

One now needs to introduce some abstract algebra. First one recalls
that the Galois field $GF(p^m)$ is the field of polynomials defined
as the quotient $\mathcal{Z}_p(x)/(q(x))$ of the ring of polynomials
$\mathcal{Z}_p(x)$ by a primitive polynomial of order $m$ over
$\mathcal{Z}_p=GF(p)$. By definition, this primitive element,
$q(x)$, has the property to be irreducible over the basic field
$GF(p)$, i.e., it cannot be factored into products of less-degree
polynomials; it is also primitive over $GF(p)$ of order $p-1$ in the
sense that it has a root $\alpha$ which generates any non zero
element of $GF(p)$ by a power sequence
$(\alpha^1,~\alpha^2,\cdots,\alpha^{p-1}=1)$ and in addition all of
its roots are in the extension field $GF(p^m)$. There is at least
one primitive polynomial for any extension field $GF(p^m)$. For
$p=2$ and $m=2$, $3$ and $4$ they are, for example, of the form
$q(x)=x^2+x+1$, $x^3+x+1$ and $x^4+x+1$, respectively.

A Galois ring $GR(4^m)$ of order $m$ is a ring of polynomials which
is an extension of $\mathcal{Z}_4$ of degree $m$ containing an
$r$-th root of unity, where $r=2^m-1$ \cite{Hammons94} \cite{Wan97}.
Let $h_2(x)\in\mathcal{Z}_2(x)$ be a primitive irreducible
polynomial of degree $m$. There is a unique monic polynomial
$h(x)\in\mathcal{Z}_4(x)$ of degree $m$ such that $h(x)=h_2(x)$(mod
2) and $h(x)$(mod 4) divides $x^r-1$. 
The polynomial $h(x)$ is the basic primitive polynomial and defines
the Galois ring $GR(4^m)=\mathcal{Z}_4(x)/(h(x))$ of cardinality
$4^m$. This ring can be found as follows. Let $h_2(x)=e(x)-d(x)$,
where $e(x)$ contains only even powers and $d(x)$ only odd powers;
then $h(x^2)=\pm(e^2(x)-d^2(x))$. For $m=2$, $3$ and $4$ one gets
$h(x)=x^2+x+1$, $x^3+2x^2+x-1$ and $x^4+2x^2-x+1$, respectively.

Any non zero element of $GF(p^m)$ can be expressed in terms of a
single primitive element. This is no longer true in $GR(4^m)$, which
contains zero divisors. But in the latter case there exists a
nonzero element $\xi$ of order $2^m-1$ which is a root of the basic
primitive polynomial $h(x)$. Any element $\beta \in GR(4^m)$ can be
uniquely determined in the form $\beta=a \oplus 2 \odot b$, where
$a$ and $b$ belong to the so-called Teichm\"{u}ller set
$\mathcal{T}_m = (0,1,\xi,\cdots,\xi^{2^m-2})$. Moreover, one finds
that $a=\beta^{2^m}$. We can also define the trace to the basis ring
$\mathcal{Z}_4$ by the map
\begin{equation}
tr(\beta)=\sum_{k=0}^{2^m-1}\sigma^k(\beta), \label{trace2}
\end{equation}
where the summation runs over the elements of the Teichm\"uller set
and the Frobenius automorphism $\sigma$ reads
\begin{equation}
\sigma(a\oplus 2 \odot b)=a^2\oplus 2\odot b^2,
\end{equation}
with $a^2 \equiv a\odot a$. Using the 2-adic decomposition of $k$ in
the exponent of (\ref{Galois}) and the above-given trace map, we
finally get
\begin{equation}
|\theta_b^a\rangle=\frac{1}{\sqrt{2^m}}\sum_{n=0}^{2^m-1}i^{tr[(a\oplus
2 \odot b)\odot n]}|n\rangle \label{evendits};
\end{equation}
the last expression gives a set of $d=2^m$ bases with index $a$ for
the basis and index $b$ for the vectors in the basis, mutually
unbiased to each other and to the computational base $B_0$
\cite{Klapp03}.

Let us apply this formula to the case of quartits. In
$GR(4^2)=\mathcal{Z}_4[x]/(x^2+x+1)$ the Teichm\"{u}ller set reads
$\mathcal{T}_2=(0,1,x,3+3x)$; the $16$ elements $a\oplus 2 \odot b$
with $a$ and $b$ in $\mathcal{T}_2$ are shown in the following
matrix
\begin{equation}\label{Teich}
 \left [\begin{array}{cccc} 0 & 2&2x&2+2x\\ 1 & 3&1+2x&3+2x\\
 x&2+x&3x&2+3x\\3+3x&1+3x&3+x&1+x \\ 
\end{array}\right].
\end{equation}
Extracting the Teichm\"{u}ller decomposition $(a\oplus 2 \odot
b)\odot n=a'\oplus 2 \odot b'$ and calculating the exponent
$tr(a'\oplus 2 \odot b')=a'\oplus 2\odot b'\oplus a'^2\oplus 2\odot
b'^2$ one gets the four MUBs
\begin{eqnarray}
&B_1=(1/2)\{(1,1,1,1),(1,1,-1,-1),(1,-1,-1,1),(1,-1,1,-1)\}\nonumber \\
&B_2=(1/2)\{(1,-1,-i,-i),(1,-1,i,i),(1,1,i,-i),(1,1,-i,i)\}\nonumber \\
&B_3=(1/2)\{(1,-i,-i,-1),(1,-i,i,1),(1,i,i,-1),(1,i,-i,1)\}\nonumber \\
&B_4=(1/2)\{(1,-i,-1,-i),(1,-i,1,i),(1,i,1,-i),(1,i,-1,i)\}.
\label{quartrits}
\end{eqnarray}
The case of $8$-dits can be examined in a similar fashion, with the
ring $GR(4^3)=\mathcal{Z}_4[x]/(x^3+2x^2+x-1)$ and Teichm\"{u}ller
set featuring the following eight elements:
$\mathcal{T}_2=\{0,1,x,x^2,1+3x+2x^2,2+3x+3x^2,3+3x+x^2,1+2x+x^2\}$.

\section{MUB'S FOR NON-PRIME-POWER DIMENSIONS} 
\label{compos}

Here we shall discuss in some detail the simplest case, which is
$d=6$, the lowest non-prime-power (n-p-p) dimension. For this case,
one constructs a set of three MUBs as follows. One takes the three
MUBs in $d=2$, viz.
\begin{equation}
B_0^{(1)}=(|0\rangle,|1\rangle),~B_1^{(1)}=(|\theta_0\rangle,|\theta_1\rangle),~B_2^{(1)}=(|\psi_0\rangle,|\psi_1\rangle),
\end{equation}
or, in the matrix form, $B_0^{(1)}$=$I_2,B_1^{(1)}$=$H$ and
$B_2^{(1)}$=$HS$, and the first three MUBs in $d=3$, viz.
\begin{equation}
B_0^{(2)}=(|0\rangle,|1\rangle,|2\rangle),~B_1^{(2)}=
(|u_0\rangle,|u_1\rangle,|u_2\rangle),~B_2^{(2)}=(|v_0\rangle,|v_1\rangle,|v_2\rangle),
\end{equation}
or, in a more convenient form 
\begin{equation}
B_0^{(2)}=I_3,\, B_1^{(2)}=\frac{1}{\sqrt{3}}\left
[\begin{array}{ccc} 1
& 1&1\\ 1 & \omega_3&\bar{\omega_3}\\
 1&\bar{\omega_3}&\omega_3\nonumber \\
\end{array}\right], \, B_2^{(2)}=\frac{1}{\sqrt{3}}\left [\begin{array}{ccc} 1 & \omega_3&\omega_3\\ 1 & \bar{\omega_3}&1\\
 1&1&\bar{\omega_3}\\ 
\end{array}\right] \quad (17a)\\
\end{equation}
and extracts the expressions for three MUBs in $d=6$ from the rows
of the following tensorial product matrices $C_0=B_0^{(1)}\otimes
B_0^{(2)}=I_6$, $C_1=B_1^{(1)}\otimes B_1^{(2)}$ and
$C_2=B_2^{(1)}\otimes B_2^{(2)}$. This construction can easily be
generalized to any n-p-p dimension \cite{Klapp03} \cite{Zauner}. One
considers the prime number decomposition $d=\prod_{i=1}^rp_i^{e_i}$,
takes its smallest factor $\tilde{m}= min_i(p_i^{e_i})$, and gets
$\tilde{m}+1$ MUBs from the tensorial product
$B^{(k)}=\otimes_{i=1}^r B_i^{(k)}$, ($k=0,..,\tilde{m}$).

At this point, it is instructive to enlighten the above-described
construction of MUBs by confining ourselves to the Galois ring in
$d=6$. Let us take the latter as the quotient
$GR(6^2)=\mathcal{Z}_6[x]/(x^2+3x+1)$ of polynomials over
$\mathcal{Z}_6$ by a polynomial irreducible over both
$\mathcal{Z}_2$ and $\mathcal{Z}_3$. $GR(6^2)$ has $36$ elements.
The notion of Teichm\"{u}ller set can be generalized to the
so-called Sylow decomposition \cite{Archer}. Any element $\beta \in
GR(6)$ can be uniquely determined in the form $\beta=a \oplus b$,
where $a$ and $b$ are in the Sylow subgroups $S_a$ and $S_b$. These
can be defined as $S_a=\{x\in GR(6): 2x=0\}$ and $S_b=\{x\in GR(6):
3x=0\}$, i.e.
\begin{eqnarray}
&S_a=\{0,3,3x,3+3x\},\nonumber \\
&S_b=\{0,2,4,2x,4x,2+2x,2+4x,4+2x,4+4x\}.
\end{eqnarray}
Since the quotient polynomial is irreducible, one notices that $S_a$
and $S_b$ themselves are finite fields, being isomorphic to $GF(4)$
and $GF(9)$, respectively. One can therefore express the ring in
dimension $6$ as $GF(4)\oplus GF(9)=GR(6)$. Can this property be
useful to construct MUBs themselves, or it merely represents a
constraint on the maximum number of MUBs? One construction of MUBs
for $d$=$6$ was based on the tensorial product of MUBs in dimension
$2$ and $3$, respectively. But the three MUBs in dimension two do
not follow from the four elements of $GF(4)$, but from the four
elements of $GR(4^1)=\mathcal{Z}_4$. On the other hand, the four
MUBs in $d$=$3$ follow from the three elements of
$GF(3)=\mathcal{Z}_3$. So the decomposition of $GR(6)$ as a product
of two fields appears to be irrelevant to the topic of MUBs.
Moreover, it was shown that complete sets of MUBs in n-p-p
dimensions cannot be constructed using a majority of generalizations
of known formulas for finite rings \cite{Archer}. This, however,
should not deter us from looking at other possible constructions.
For example, using the properties of sets of mutually orthogonal
Latin squares, it has recently been shown that in the particular
square dimension $d=26^2$ it is, in principle, possible to construct
at least $6$ MUBs, while the construction based on the prime number
decomposition determines only $min_i(p_i^{e_i})+1=2^2+1=5$ of them
\cite{Wocjan04}.

\section{ MUB'S AND FINITE PROJECTIVE PLANES} 
\label{projective}

An intriguing similarity between mutually unbiased measurements and
finite projective geometry has recently been noticed \cite{Saniga}.
Let us find the minimum number of different measurements we need to
determine uniquely the state of an ensemble of identical $d$-state
particles. The density matrix of such an ensemble, being Hermitic
and of unit trace, is specified by $(2d^2/2)-1=d^2-1$ real
parameters. When one performs a non-degenerate orthogonal
measurement on each of many copies of such a system one eventually
obtains $d-1$ real numbers (the probabilities of all but one of the
$d$ possible outcomes). The minimum number of different measurements
needed to determine the state uniquely is thus $(d^2-1)/(d-1)= d+1$
\cite{Wootters89} \cite{Wootters04bis}.

It is striking that the identical expression can be found within the
context of finite projective geometry. A finite projective plane is
an incidence structure consisting of points and lines such that any
two points lie on just one line, any two lines pass through just one
point, and there exist four points, no three of them on a line
\cite{Beutel98}. From these properties it readily follows that for
any finite projective plane there exists an integer $d$ with the
properties that any line contains exactly $d+1$ points, any point is
the intersection of exactly $d+1$ lines, and the number of points is
the same as the number of lines, namely $d^{2}+d+1$. This integer
$d$ is called the order of the projective plane. The most striking
issue here is that the order of known finite projective planes is a
power of prime. The question of which other integers occur as orders
of finite projective planes remains one of the most challenging
problems of contemporary mathematics. The only ``no-go" theorem
known so far in this respect is the Bruck-Ryser theorem
\cite{Bruck49} saying that there is no projective plane of order $d$
if $d-1$ or $d-2$ is divisible by 4 and $d$ is not the sum of two
squares. Out of the first few non-prime-power numbers, this theorem
rules out finite projective planes of order 6, 14, 21, 22, 30 and
33. Moreover, using massive computer calculations, it was proved
that there is no projective plane of order ten. It is surmised that
the order of any projective plane is a power of a prime.

It has been conjectured by Saniga and two of us \cite{Saniga} that
the question of the existence of a set of $d+1$ MUBs in a
$d$-dimensional Hilbert space if $d$ differs from a power of a prime
number is identical with the problem of whether there exist
projective planes whose order $d$ is not a power of a prime number.
Furthermore, for power of a prime dimension, the complete sets of
MUBs can be put in correspondence with $d+1$-arcs, which are
`curves' known as {\em ovals} in (Desarguesian) projective plane of
order $d$ \cite{sp05}. For $d=2^n$ and $n\geq 3$ there are two types
of ovals, viz. conics and non-conics, implying the existence of two
types of MUBs for such dimensions. In addition, in the same case of
a power of a prime dimension $d=p^r$, the $p^r$ vectors of a basis
correspond to the total number of points in a so-called neighbour
class along a (proper) conic of a projective Hjelmslev plane defined
over a Galois ring of characteristic $p^2$ and rank $r$, whereas the
$d+1$ MUBs correspond exactly to the total number of pairwise
disjoint neighbour classes on the conic \cite{spH}.

\subsection{ $GF(8)$ and the Fano plane}  

The smallest projective plane, also called the Fano plane, is
obviously the $d=2$ one; it contains $7$ points and $7$ lines, any
line contains $3$ points and each point is on $3$ lines. It
comprises a $3$-dimensional vector space over the field $GF(2)$,
each point being a triple $(g_1,g_2,g_3)$, excluding the (0,0,0)
one, where $g_i \in GF(2) = \{0,1\}$ \cite{Beutel98}. The points of
this plane can also be represented in terms of the non-zero elements
of the Galois field $G=GF(2^3)$.

To see this, we recall that this field is isomorphic to
$\mathcal{Z}_2(x)/(p(x))$ with the polynomial $p(x)=x^3+x+1$
irreducible in $GF(2)$. It is well-known that there are three useful
representations of the elements of $GF(8)$ as shown in Table 1
\cite{Beutel98} \cite{Hirsch98} \cite{Batten97}.

\begin{table}[htbp]
\centering \caption{Representations of the elements of the Galois
field $GF(8)$ \label{Representations of the elements of the Galois
field $GF(8)$}}

\begin{tabular}{|c|c|c|}
\hline
 as powers of $\alpha$ &as polynomials&as $3$-tuples in $\mathcal{Z}_2^3$\\
\hline \hline
0 & 0 &(0,0,0)\\
\hline
1&1&(0,0,1)\\
\hline
$\alpha$ & $x$ & (0,1,0)\\
\hline
$\alpha^2$&$x^2$&(1,0,0)\\
\hline
$\alpha^3$&$1+x$&(0,1,1)\\
\hline
$\alpha^4$&$x+x^2$&(1,1,0)\\
\hline
$\alpha^5$&1+$x$+$x^2$&(1,1,1)\\
\hline
$\alpha^6$&$1+x^2$&(1,0,1)\\
\hline
\end{tabular}
\end{table}

The first representation emphasizes the fact that $G^*=G-\{0\}$ is a
multiplicative cyclic group of order $7$, for $\alpha^7=1$. The
second representation is obtained from the first by calculating
modulo the primitive polynomial $p(x)$. Finally, the $3$-tuple
representation is obtained from the coefficients of the three powers
$x^0=1$, $x^1=x$ and $x^2$. Taking these $3$-tuples as the points of
a $3$-dimensional vector space, we recover the Fano plane $-$ see
Fig.~1.

\subsection{The lifted Fano plane in $GR(4^3)$}

We already know from Sect. \ref{MUBsQubits} that the relevant object
for $2^m$-dits is not the Galois field $GF(2^m)$, but rather the
Galois ring $GR(4^m)$. It is therefore important to have a look at
the geometry in the space $A=GR(4^3)$. For a ring, the concept of a
vector space must be replaced by that of a module. The largest cycle
in $A$ is the set $\mathcal{T}_3^*=\mathcal{T}_3-\{0\}$ (see Sect.
3), and each element of $\mathcal{T}_3^*$ can be represented in the
same way as in the case of a Galois field. This is summarized in
Table 2.
\begin{table}[htbp]
\centering \caption{Representations of the elements of the cyclic
group in the Galois ring $GR(4^3)$ \label{Representations of the
elements of the cyclic group in the Galois ring $GR(4^3)$}}
\begin{tabular}{|c|c|c|c|}
\hline
 as powers of $\xi$ &as polynomials&as $3$-tuples in $\mathcal{Z}_4^3$&as $3$-tuples in $\mathcal{Z}_2^3$\\
\hline \hline
0&0&(0,0,0)&(0,0,0)\\
\hline
1&1&(0,0,1)&(0,0,1)\\
\hline
$\xi$ & $x$ & (0,1,0)&(0,1,0)\\
\hline
$\xi^2$&$x^2$&(1,0,0)&(1,0,0)\\
\hline
$\xi^3$&$1+3x+2x^2$&(2,3,1)&(0,1,1)\\
\hline
$\xi^4$&$2+3x+3x^2$&(3,3,2)&(1,1,0)\\
\hline
$\xi^5$&$3+3x$+$x^2$&(1,3,3)&(1,1,1)\\
\hline
$\xi^6$&$1+2x+x^2$&(1,2,1)&(1,0,1)\\
\hline
\end{tabular}
\end{table}
Any polynomial $h(x)$ in $\mathcal{T}_3^*$ (column 2) is uniquely
projected as a polynomial $h_2(x)=h(x)$ (mod 2) in $GF(8)$, which
results in the $3$-tuple representation in $\mathcal{Z}_2^3$ (column
4). Vice versa, any polynomial in $GF(8)$ has a unique lift in
$\mathcal{T}_3^*$. Since the geometrical structure we are looking at
is combinatorial and doesn't depend on particular coordinates, it
follows that the lifted Fano plane in $\mathcal{T}_3^*$ is still the
Fano plane up to isomorphism. So the Fano geometry is inherent in
the geometry of qubits, but we needed a special coordinatization in
order to be able to see that. The very recent work of Planat,
Saniga, and Kibler \cite{psk} relates the understanding of MUBs, the
Fano plane, entanglement, and quantum paradoxes to the construction
of projective lines over special families of finite rings.

\begin{figure}[htb]
\centerline{
\includegraphics[width=7.15cm]{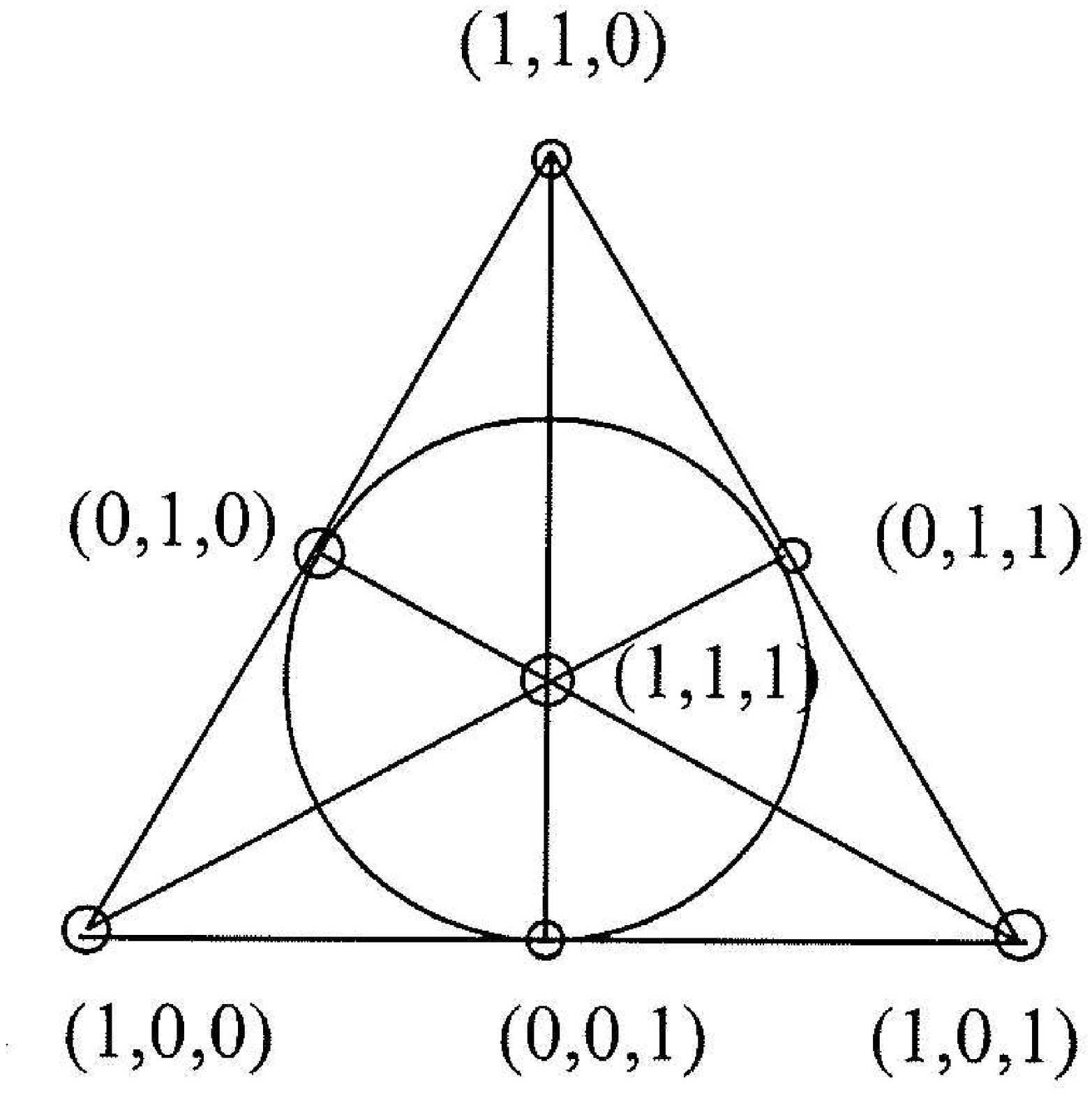}}
\caption{The Fano plane. } \label{RFTAGN3}
\end{figure}

\section{ORTHOGONAL SETS OF MAXIMALLY\\ ENTANGLED STATES}
\label{seis}

The above-discussed methods of constructing MUBs can
straightforwardly be used for recognizing orthogonal partial bases
of maximally entangled states, of which some can be mutually
unbiased in their corresponding subspaces implying that the MUB
normalization $1/\sqrt{d}$ is replaced by a higher normalization
$1/\sqrt{d_s}$ with $d_s < d$. In this section, by partial bases we
mean orthogonal sets of Hilbert vectors that are not in enough
number to form complete bases. Following the methodology outlined in
Sections \ref{MUBGaloisFields} and \ref{MUBsQubits}, let us consider
a set of generalized Bell states defined as a two particle quantum
Fourier transform \cite{Cerf01} \cite{Fivel95}
\begin{equation}
|\mathcal{B}_{h,k}\rangle=\frac{1}{\sqrt{d}}\sum_{n=0}^{d-1}\omega_d^{k
n}|n,n+h\rangle, \label{FourierEntang}
\end{equation}
where $|n,n+h\rangle$ denotes the two-particle state
$|n\rangle|n+h\rangle$ and the operation $n+h$ is performed modulo
$d$. These states are both orthonormal, $\langle
\mathcal{B}_{h,k}|\mathcal{B}_{h',k'} \rangle
=\delta_{hh'}\delta_{kk'}$, and maximally entangled,
$trace_2|\mathcal{B}_{h,k}\rangle \langle\mathcal{B}_{h,k}|
=\frac{1}{d}I_d$, where $trace_2$ means the partial trace over the
second qudit \cite{Nielsen00}. If one restricts to the case of
$2$-qubits, one recovers the well-known representation of Bell
states
%
$(|\mathcal{B}_{0,0}\rangle,|\mathcal{B}_{0,1}\rangle)=\frac{1}{\sqrt{2}}(|00\rangle+|11\rangle,|00\rangle-|11\rangle),
(|\mathcal{B}_{1,0}\rangle,|\mathcal{B}_{1,1}\rangle)$=

\noindent $
\frac{1}{\sqrt{2}}(|01\rangle+|10\rangle,|01\rangle-|10\rangle)$,
%
where a more compact notation $|00\rangle=|0,0\rangle$,
$|01\rangle=|0,1\rangle$,\dots, is employed. Let us first focus on
the $2$-qubit case. 
Paralleling what we did in Section \ref{MUBsQubits}, $kn$ in
(\ref{FourierEntang}) is first identified as the multiplication
$k\odot n$ of polynomials in $GR(4)$ and then $k$ is Teichm\"{u}ller
decomposed, i.e., $k=a \oplus 2\odot b $. This leads to $4$ sets
($h, a=0,1$) of two vectors $(b=0,1)$, namely
\begin{equation}
|\mathcal{B}_{h,b}^a\rangle=\frac{1}{\sqrt{2}}\sum_{n=0}^1i^{(a\oplus
2 \odot b)\odot n}|n,n\oplus h\rangle. \label{twoqubits}
\end{equation}
Casting the last equation into its matrix form (save for the
proportionality factor),
\begin{equation}
 \left [\begin{array}{cc} (|00\rangle)+|11\rangle,|00\rangle-|11\rangle);&(|01\rangle+|10\rangle,|01\rangle-|10\rangle)
 \\(|00\rangle+i|11\rangle,|00\rangle-i|11\rangle);&(|01\rangle+i|10\rangle,|01\rangle)-i|10\rangle)\\
\end{array}\right],
\end{equation}
one finds that two partial bases in one column are mutually unbiased
in their subspace, while the vectors in two partial bases on the
same line are orthogonal to each other.

Eq.~(\ref{twoqubits}) can easily be extended to maximally entangled
two-particle sets of $2^m$-dits by applying, as in
Eq.~(\ref{evendits}), the Frobenius map (\ref{trace2}) to the basis
field $\mathcal{Z}_4$
\begin{equation}
|\mathcal{B}_{h,b}^a\rangle=\frac{1}{\sqrt{2^m}}\sum_{n=0}^{2^m-1}i^{tr[(a\oplus
2 \odot b)\odot n]}|n,n\oplus h\rangle \label{twoevendits}.
\end{equation}
For $2$-particle sets of quartits, using Eqs.~(\ref{quartrits}) and
(\ref{twoevendits}), one thus gets $4$ sets
$(|\mathcal{B}_{h,b}^a\rangle$, $ h=0,...,3)$ of $4$ mutually
unbiased partial bases $(a=0,...,3)$,
\begin{eqnarray}
&\{(|00\rangle+|11\rangle+|22\rangle+|33\rangle,|00\rangle+|11\rangle-|22\rangle-|33\rangle,\nonumber
\\
&|00\rangle-|11\rangle-|22\rangle+|33\rangle,|00\rangle-|11\rangle+|22\rangle-|33\rangle);\nonumber
\\
&(|00\rangle-|11\rangle-i|22\rangle-i|33\rangle,|00\rangle-|11\rangle+i|22\rangle+i|33\rangle,\nonumber\\
&|00\rangle+|11\rangle+i|22\rangle-i|33\rangle,|00\rangle+|11\rangle-i|22\rangle+i|33\rangle)
; \nonumber \\
&\cdots\nonumber\}
\end{eqnarray}
\begin{eqnarray}
&\{(|01\rangle+|12\rangle+|23\rangle+|30\rangle,|01\rangle+|12\rangle-|23\rangle-|30\rangle,\nonumber\\
&|01\rangle-|12\rangle-|23\rangle+|30\rangle,|01\rangle-|12\rangle+|23\rangle-|30\rangle);\nonumber\\
&(|01\rangle-|12\rangle-i|23\rangle-i|30\rangle,|01\rangle-|12\rangle+i|23\rangle+i|30\rangle,\nonumber\\
&|01\rangle+|12\rangle+i|23\rangle-i|30\rangle,|01\rangle+|12\rangle-i|23\rangle+i|30\rangle);\nonumber\\
&\cdots \nonumber\}
\end{eqnarray}
\begin{eqnarray}
&\{(|02\rangle+|13\rangle+|20\rangle+|31\rangle,|02\rangle+|13\rangle-|20\rangle-|31\rangle,\nonumber\\
&|02\rangle-|13\rangle-|20\rangle+|31\rangle,|02\rangle-|13\rangle+|20\rangle-|31\rangle);\cdots\nonumber\\
&\cdots\nonumber\}
\end{eqnarray}
\begin{eqnarray}
&\{(|03\rangle+|10\rangle+|21\rangle+|32\rangle,|03\rangle+|10\rangle-|21\rangle-|32\rangle,
\nonumber\\
&|03\rangle-|10\rangle-|21\rangle+|32\rangle,|03\rangle-|10\rangle+|21\rangle-|32\rangle);
\cdots\nonumber\\
&\cdots\},
\end{eqnarray}
where we have skipped the partial normalization factor ($1/2$).
Within each set, the four partial bases are mutually unbiased, as in
(\ref{quartrits}), while the vectors of the partial bases from
different sets are orthogonal.

Turning now to odd characteristic, i.e. to $d=p^m$ with $p$ an odd
prime, we can similarly extend Wootters formula (\ref{newGalois}) to
the generalized Bell states
\begin{equation}
|\mathcal{B}_{h,b}^a\rangle=\frac{1}{\sqrt{d}}\sum_{n=0}^{d-1}\omega_d^{tr[(a
\odot n \oplus b)\odot n ]}|n,n\oplus h\rangle,
\label{entangledGalois}
\end{equation}
where the trace is defined by Eq.~(\ref{trace}). A list of the
generalized Bell states of qutrits for $a=0$ can be found in
\cite{Fujii01}, a work that relies on a coherent state formulation
of entanglement. In general, for $d$ a power of a prime, starting
from (\ref{FourierEntang}) or (\ref{entangledGalois}) one obtains
$d^2$ sets of $d$ maximally entangled states. Each set of the $d$
bases (with $h$ fixed) has the property of mutual
unbiasedness.\\

Eq.~(\ref{FourierEntang}) can be used, without any substantial
restriction, to find $d$ sets ($h=0,..,d-1$) of maximally entangled
states in any composite dimension $d=\prod_{i=1}^rp_i^{e_i}$. Or one
can also follow the strategy of Section \ref{compos} to get
$\tilde{m}=min_i(p_i^{e_i})$ sets of MUBs of maximally entangled
states. In $d=6$, for example, one expects that two such $d$ sets
can be constructed. Using the tensorial products in Sect.
\ref{compos}, one indeed finds the two $2\times6$ sets (with the
$1/\sqrt{6}$ factor omitted) {\tiny \flushleft
\begin{eqnarray}
& \{(|00\rangle + |11\rangle + |22\rangle +|33\rangle + |44\rangle +
|55\rangle, |00\rangle + \omega_3|11\rangle +
\bar{\omega_3}|22\rangle +|33\rangle + \omega_3|44\rangle +
\bar{\omega_3}|55\rangle,\nonumber \\
& |00\rangle + \bar{\omega_3}|11\rangle + \omega_3|22\rangle
+|33\rangle + \bar{\omega_3}|44\rangle+ \omega_3|55\rangle\nonumber,
|00\rangle + |11\rangle + |22\rangle
-|33\rangle - |44\rangle - |55\rangle,\nonumber\\
& |00\rangle + \omega_3|11\rangle + \bar{\omega_3}|22\rangle
-|33\rangle - \omega_3|44\rangle - \bar{\omega_3}|55\rangle,
|00\rangle + \bar{\omega_3}|11\rangle + \omega_3|22\rangle
-|33\rangle - \bar{\omega_3}|44\rangle -
\omega_3|55\rangle);\nonumber\\
& (|00\rangle + \omega_3|11\rangle + \omega_3|22\rangle +i|33\rangle
+ i\omega_3|44\rangle + i\omega_3|55\rangle, |00\rangle +
\bar{\omega_3}|11\rangle + |22\rangle +i|33\rangle +
i\bar{\omega_3}|44\rangle +
i|55\rangle,\nonumber \\
& |00\rangle + |11\rangle + \bar{\omega_3}|22\rangle +i|33\rangle+
i|44\rangle +i\bar{\omega_3} |55\rangle,|00\rangle +
\omega_3|11\rangle + \omega_3|22\rangle -i|33\rangle -
i\omega_3|44\rangle -
i\omega_3|55\rangle,\nonumber \\
 &
|00\rangle + \bar{\omega_3}|11\rangle + |22\rangle -i|33\rangle
- i\bar{\omega_3}|44\rangle - i|55\rangle, |00\rangle + |11\rangle +
\bar{\omega_3}|22\rangle -i|33\rangle- i|44\rangle -i\bar{\omega_3}
|55\rangle);\ldots\}\nonumber
\end{eqnarray}
\begin{eqnarray}
&.\nonumber\\
&.\nonumber\\&.\nonumber\\
&\{(|01\rangle + |12\rangle + |23\rangle +|34\rangle + |45\rangle +
|50\rangle, |01\rangle + \omega_3|12\rangle +
\bar{\omega_3}|23\rangle +|34\rangle + \omega_3|45\rangle +
\bar{\omega_3}|50\rangle,\cdots\}.
\end{eqnarray}}
\normalsize

Multipartite entanglement is a key ingredient of many quantum
protocols, still needing much work to be properly understood. Sets
of orthogonal product states that are not extendible, meaning that
no further product states can be found orthogonal to all the
existing ones, have recently attracted a lot of attention. These
non-extendible product bases \cite{Vincenzo03}, and their complement
\cite{Horodecki97}, certainly deserve reconsideration in terms of
the above-outlined theory, which is based on abstract algebra and
finite geometry.

The Fourier transform approach implies that mutual unbiasedness and
maximal entanglement are complementary aspects in orthogonal quantum
measurements. In such measurements, the quantum states are encoded
in a three-dimensional lattice of indices $h$ (entanglement), $a$
(unbiasedness) and $b$ (dimensionality of Hilbert space). If $d$ is
a power of a prime, the lattice is a cube since in this case $h$,
$a$ and $b$ reach their limiting value $d$. If one forgets about
entanglement ($h=0$), the finite geometry which seems to be of most
relevance is that of a finite projective plane. On the other hand,
when unbiasedness is not taken into account, as well as for
multipartite information tasks when $d$ is not (a power of) a prime,
other concepts have been introduced, such as Bell inequalities
\cite{Yu02}, coherent states \cite{Fujii01}, entanglement swapping
\cite{Bose98}, generalized Hopf fibrations \cite{Bernevig03},
topological entanglement \cite{Kauffman02} and bound entanglement
\cite{Vincenzo03}, to mention a few.


\section*{ACKNOWLEDGEMENTS}

\noindent 
H.C.R. acknowledges partial support from the
Mexican CONACyT project 46980.










\begin{thebibliography}{99}



\bibitem {Quiprocone} Quiprocone website,
http://www.imaph.tu-bs.de/qi/problems
\bibitem{s60} J. Schwinger, ``Unitary operator bases, {\em Proc.
Nat. Acad. Sci. U.S.A.} {\bf 46}, 560 (1960).
\bibitem{a80} W.O. Alltop, ``Complex sequences with low periodic
correlations", {\em IEEE Transactions on Inf. Th.} {\bf 26}, 350
(1980).
\bibitem{iv81} I.D. Ivanovi\'c, ``Geometrical description of quantal
state determination", {\em J. Phys. A} {\bf 14}, 3241 (1981).
\bibitem{Wootters89} W.K. Wootters and B.D. Fields,
``Optimal state-determination by mutually unbiased measurements",
{\it Ann. Phys. (N.Y.)} \textbf{191}, 363 (1989).
\bibitem{Archer} C. Archer, ``There is no generalization of known formulas for MUBs",
{\it J. Math. Phys.} {\bf 46}, 022106 (2005).
\bibitem{Klapp03}A. Klappenecker and M. R\"{o}tteler,
``Construction of MUBs", {\it Lect. Notes in Comp. Science} {\bf
2948}, 137 (2004).
\bibitem{Wocjan04} P. Wocjan and T. Beth, ``New construction of MUBs in square dimensions",
{\it Quant. Inf. Comput.} {\bf 5}, 181 (2005).
\bibitem{Grassl} M. Grassl, ``On SIC-POVMs and MUBs in dimension 6",
 {\it Proc. ERATO Conf. on Quant. Inf. Science (EQUIS 2004)}, pp. 60-61 (2004). 
\bibitem{Wootters04} W.K. Wootters, ``Quantum measurements and finite geometries",
{\it Found. Phys.} {\bf 36}, 112 (2006). 
\bibitem{comb06} M. Combescure, ``The MUBs revisited", preprint
quant-ph/0605090 (2006).
\bibitem{Merolla99} J.M. Merolla, Y. Mazurenko, J.P. Goedgebuer and
W.T. Rhodes, ``Single-photon interference in sidebands of
phase-modulated light for MUBs", {\it Phys. Rev. Lett.} \textbf{82},
1656 (1999).
\bibitem{Nielsen00} M.A. Nielsen and I. Chuang, \textit{Quantum Computation and Quantum
Information} (Cambridge University Press, Cambridge, 2000), p. 582.
\bibitem{Cerf01} N.J. Cerf, M. Bourennane, A. Karlsson and N.
Gisin, ``Security of quantum key distribution using d-level
systems", {\it Phys. Rev. Lett.} \textbf{88}, 127902 (2002).
\bibitem{Durt03} T. Durt, D. Kaszlikowski, J.L. Chen, L.C. Kwek,
``Security of quantum key distribution with entangled qudits", {\em
Phys.
Rev. A} {\bf 69}, 032313 (2004).
\bibitem{Renes03} J.M. Renes, R. Blume-Kohout, A.J. Scott and C.M. Caves,
``Symmetric informationally complete quantum measurements", {\it J.
Math. Phys.} {\bf 45}, 2171 (2004).
\bibitem{Saniga} M. Saniga, M. Planat and H. Rosu, ``MUBs and finite projective planes",
{\it J. Opt. B: Quantum Semiclass. Opt.} \textbf{6}, L19 (2004).
\bibitem{be05} I. Bengtsson and A. Ericsson, ``MUBs and the
complementarity polytope", {\it Open Sys. \& Inf.} {\bf 12}, 107
(2005).
\bibitem{boy05} P.O. Boykin, M. Sitharam, P.H. Tiep, P. Wocjan,
``MUBs and orthogonal decompositions of Lie algebras", preprint
quant-ph/0506089 (2005). 
\bibitem{Kib1} M.R. Kibler, ``Angular momentum and MUBs", {\it Int. J. Mod. Phys. B} {\bf 20}, 1792
(2006).
\bibitem{Kib2} M.R. Kibler and M. Planat, ``A SU(2) recipe for MUBs", {\it Int. J. Mod. Phys. B} {\bf 20}, 1802
(2006).
\bibitem{hor3} M. Horodecki, P. Horodecki and R. Horodecki,
``Mixed-state entanglement and distillation: Is there a ``bound"
entanglement in nature ?", {\it Phys. Rev. Lett.} {\bf 80}, 5239
(1998); B. Baumgartner, B.C. Hiesmayr and H. Narnhofer, ``The state
space for 2 qutrits has a phase space structure in its core",
preprint quant-ph/0606083 (2006).
\bibitem{rom05} J.L. Romero, G. Bj\"ork, A.B. Klimov and L.L.
S\'anchez-Soto, ``Structure of the sets of MUBs for $N$ qubits",
{\it Phys. Rev. A} {\bf 72}, 062310 (2005).
\bibitem{Vourdas04} A. Vourdas, ``Quantum systems with finite Hilbert space",
{\it Rep. Prog. Phys.} \textbf{67}, 267 (2004); ``The Frobenius
formalism in Galois quantum systems", preprint quant-ph/0605054
(2006).
\bibitem{Planat04} M. Planat and H. Rosu, ``The hyperbolic, the arithmetic and the quantum phase",
{\it J. Opt. B: Quantum Semiclass. Opt.} \textbf{6}, S583 (2004);
``Mutually unbiased phase states, phase uncertainties, and Gauss
sums", {\it Eur. Phys. J. D} {\bf 36}, 133 (2005).
\bibitem{Bandyo01} S. Bandyopadhyay, P.O. Boykin, V. Roychowdhury
and F. Vatan, ``A new proof for the existence of MUBs", {\it
Algorithmica} \textbf{34}, 512 (2002).
\bibitem{Lidl}R. Lidl and G. Pilz, \textit{Applied Abstract
Algebra}, Second Edition (Springer, New York, 1998).
\bibitem{Durt04} T. Durt, ``If 1=2+3, then 1=2$\cdot$3: Bell states, finite groups and MUBs, a unifying approach",
preprint quant-ph/0401046 (2004).
\bibitem{Partha04} K.R. Parthasarathy, ``On estimating the state of a finite level quantum system",
preprint quant-ph/0408069 (2004).
\bibitem{Wootters04bis} K.S. Gibbons, M.J. Hoffman and W.K.
Wootters, ``Discrete phase space based on finite fields", {\it Phys.
Rev. A} {\bf 70}, 062101 (2004). 
\bibitem{Mosseri03}C. Rigetti, R. Mosseri and M. Devoret,
``Geometric approach to digital quantum information", {\it Quant.
Inf. Processing} {\bf 3}, 351 (2004). 
\bibitem{Pittenger03}A.O. Pittenger and M.H. Rubin, ``MUBs,
generalized spin matrices and separability",
{\it Linear Alg. Appl.} {\bf 390}, 255 (2004).
\bibitem{Hammons94} A.R. Hammons, P.V. Kumar, A.R. Calderbank,
N.J.A. Sloane and P. Sol\'e, ``The $Z_4$-linearity of Kerdock,
Preparata, Goethals, and related codes", {\it IEEE Trans. Inform.
Theory} \textbf{40}, 301 (1994).
\bibitem{Wan97} Z.X. Wan, \textit{Quaternary Codes} (World
Scientific, Singapore, 1997).
\bibitem{Zauner} G. Zauner, \textit{Quantendesigns-Grundz\"{u}ge einer nichtkommutativen Designtheorie}
 (Dissertation, Universit\"{a}t Wien, 1999).
\bibitem{Beutel98} A. Beutelspacher and U. Rosenbaum, \textit{Projective
geometry: from foundations to applications} (Cambridge University
Press, Cambridge, 1998).
\bibitem{Bruck49} R.H. Bruck and H.J. Ryser, ``The nonexistence of certain finite projective planes",
{\it Can. J. Math.} \textbf{1}, 88 (1949).
\bibitem{sp05} M. Saniga and M. Planat, ``Sets of MUBs as arcs in
finite projective planes ?", {\em Chaos, Solitos and Fractals} {\bf
26}, 1267 (2005).
\bibitem{spH} M. Saniga and M. Planat, ``Hjelmslev geometry of
MUBs", {\em J. Phys. A} {\bf 39}, 435 (2006).
 \bibitem{Hirsch98}J.W.P. Hirschfeld, \textit{Projective geometries over finite
fields} (Oxford University Press, Oxford, 1998).
\bibitem{Batten97} L.M. Batten, \textit{Combinatorics of finite
geometries} (Cambridge University Press, Cambridge, 1997).
\bibitem{psk} M. Planat, M. Saniga and M.R. Kibler, ``Quantum
entanglement and projective ring geometry", {\it SIGMA} \textbf{2},
paper 066 (2006).
\bibitem{Fivel95} D.I. Fivel, ``Remarkable phase oscillations appearing in the lattice dynamics of EPR states",
{\it Phys. Rev. Lett.} \textbf{74}, 835 (1995).
\bibitem{Fujii01} K. Fujii, ``A relation between coherent states and generalized Bell states",
preprint quant-ph/0105077 (2001).
\bibitem{Vincenzo03} D.P. DiVincenzo, T. Mor, P.W. Shor, J.A.
Smolin and B.M. Terhal, ``Unextendible product bases, uncompletable
product bases and bound entanglement",  {\it Commun. Math. Phys.}
\textbf{238}, 379 (2003).
\bibitem{Horodecki97} P. Horodecki, ``Separability criterion and inseparable mixed states with positive partial transposition",
{\it Phys. Lett. A} \textbf{232}, 333 (1997).
\bibitem{Yu02} S. Yu, Z.B. Chen , J.W. Pan and Y.D. Zhang,
``Classifying N-qubit entanglement via Bell's inequalities", {\it
Phys. Rev.
Lett.} {\bf 90}, 080401 (2003).
\bibitem{Bose98} S. Bose, V. Vedral and P.L. Knight,
``Multiparticle generalization of entanglement swapping", {\it Phys.
Rev. A} \textbf{57}, 822 (1998).
\bibitem{Bernevig03} B.A. Bernevig and H.D. Chen,
``Geometry of the 3-qubit state, entanglement and division
algebras",
{\it J. Phys. A} \textbf{36}, 8325 (2003).
\bibitem{Kauffman02} L.H. Kauffman and S.J. Lomonaco,
``Quantum entanglement and topological entanglement", {\it New J.
Phys.} \textbf{4}, 73.1 (2002).




\end{thebibliography}
\end{document}